\documentclass[aps,prl,reprint,superscriptaddress,twocolumn]{revtex4-1}

\usepackage{graphicx}

\begin{document}

\title{Quantum dot resonant-fluorescence linewidth narrowing and enhanced intensity}

\author{O. Gazzano}
\email{Corresponding author: ogazzano@umd.edu}
\affiliation{Joint Quantum Institute, National Institute of Standards and Technology, \& University of Maryland, Gaithersburg, MD, USA.}
\author{T. Huber}
\affiliation{Joint Quantum Institute, National Institute of Standards and Technology, \& University of Maryland, Gaithersburg, MD, USA.}
\author{V. Loo}
\affiliation{Joint Quantum Institute, National Institute of Standards and Technology, \& University of Maryland, Gaithersburg, MD, USA.}
\author{S. Polyakov}
\affiliation{National Institute of Standards and Technology Gaithersburg, MD, USA.}
\author{E. B. Flagg}
\affiliation{Department of Physics and Astronomy, West Virginia University, Morgantown, WV USA.}
\author{G. S. Solomon}
\affiliation{Joint Quantum Institute, National Institute of Standards and Technology, \& University of Maryland, Gaithersburg, MD, USA.}
\affiliation{National Institute of Standards and Technology Gaithersburg, MD, USA.}

\date{\today}

\begin{abstract}
We study the processes in a quantum dot and its surrounds under resonant excitation with the addition of weak non-resonant light. We observe a decrease in inhomogeneous emission linewidth, as well as previously observed enhancement of resonant fluorescence (RF). Scanning excitation and detection frequencies, we distinguish between homogeneous and inhomogeneous broadening. Monte Carlo simulations model the enhanced RF only if we include charge-carrier loss from the quantum dot induced by the resonant laser. The cause of the linewidth narrowing is predominantly independent of the enhanced RF.
\end{abstract}

\maketitle

For many quantum information and quantum optics applications, indistinguishable single photons are necessary~\cite{Shih1988,Cirac1997,Knill2001a,Broome2013b,Carolan2015}.
Resonant excitation eliminates the time uncertainty in the state preparation and thus can improve photon indistinguishability~\cite{Kiraz2004,Flagg2012a,Huber2016}.
Semiconductor quantum dots (QDs) of InAs embedded in a solid-state host are bright emitters of single photons~\cite{Claudon2010,Gazzano2013,Gazzano2016}. In particular, resonant excitation has been shown to be essential for highly indistinguishable single photons from QD states~\cite{He2013,Kuhlmann2015,Somaschi2015}. However, in QD resonant excitation the fluorescence is often strongly reduced and remains spectrally broad, limiting the scope of applications~\cite{Metcalfe2010,Nguyen2012}. Furthermore, the photon indistinguishability drops for photons emitted temporally more than a few hundreds of nanosecond apart~\cite{Thoma2016,Loredo2016,Wang2016}. Thus for a variety of quantum information processes using QDs\textemdash for instance, entanglement distribution \cite{Cirac1997,Yin2012}, photon multiplexing~\cite{Loredo2016,Xiong2016} or multiple sources experiments~\cite{Flagg2010,Patel2010}\textemdash resonant fluorescence intensity and inhomogeneous broadening must be better understood.

\begin{figure*}
\includegraphics[trim=0cm 0cm 0cm 0cm, clip=true,width=1\linewidth]{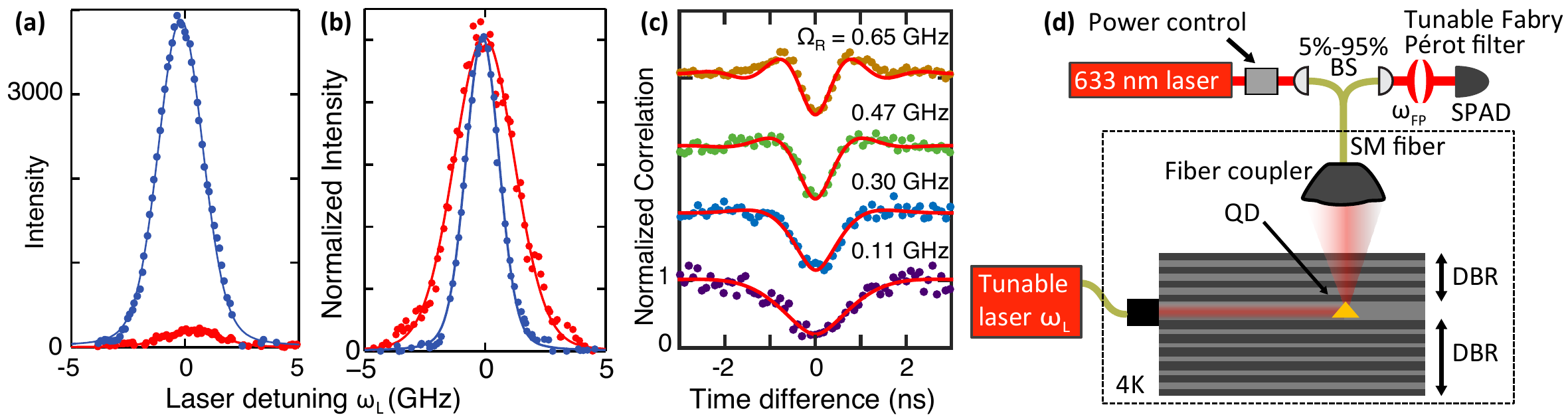}
\caption{(a-b) Resonant fluorescence excitation (PLE) spectroscopy of a QD transition for two above band 633~nm laser powers (red dots: 1~nW ; blue dots: 40~nW). The intensity is not normalized in (a) to show the strong brightness improvement by a factor of~30. The intensity is normalized in (b) to show the inhomogeneous linewidth narrowing (factor $\approx 2$). (c) Autocorrelation measurements performed under CW resonant excitation and for different Rabi frequencies (i.e. resonant laser powers). (d) Schematic of the experiment. A single-mode (SM) optical fiber is glued to the cleaved edge of the sample to resonantly excite a QD embedded between two distributed Bragg reflectors (DBR). The light is collected from the sample top surface using a fiber-coupled microscope objective. The weak, 633~nm laser is sent to the sample using the same objective. The Fabry P\'erot filter is bypassed in the PLE scans.}
\label{fig:Schematics}
\end{figure*}

Ideally, the light spontaneously emitted by an isolated two-level system has a Lorentzian line shape representing the Fourier transformation of the spontaneous emission process. The spectral linewidth, $\gamma$ is then determined by only the dipole radiative decay rate, $\Gamma$ by the relation $\gamma=\Gamma/2$. A real world solid-state emitter, however, often interacts with its environment. With environmental coupling, the emitter spectrum can be broadened and its line shape will be modified depending on the characteristic time and interaction rate of the coupling. Fast coupling mechanisms, on the timescale of $\Gamma^{-1}$, homogeneously broaden the Lorentzian line shape spectrum so that $\gamma=\Gamma/2+\gamma^{\star}$, where $\gamma^{\star}$ is the pure dephasing rate. Slow environmental coupling induces spectral wandering, and leads to an inhomogeneous lineshape. In each case the broadening leads to reduced indistinguishability.

In this letter, we demonstrate a resonant-fluorescence scheme to reduce the inhomogeneous linewidth of QD transitions and to improve the source brightness. Furthermore, we elucidate the underlying processes responsible for these effects. We show that light from a weak above band laser can improve the brightness of a resonantly pumped QD transition by a factor larger than~$30$ and reduce its inhomogeneous transition linewidth by a factor of about~2 as shown in Figs.~1.(a)-(b). The additional incoherent pump produces no measurable photons in the region of interest when the resonant laser is switched off. The brightness improvement has been previously reported~\cite{Metcalfe2010,Nguyen2012,Chen2016}, but to our knowledge, the linewidth narrowing has not. While the peak intensity increases and the inhomogeneous linewidth narrows with 
increasing above-band laser intensity, we show that the intensity integrated over a sweep of the resonant laser frequency is not constant. Simple Monte Carlo simulations with minimal assumptions show that carrier dynamics inside the QD can be stabilized by the additional above-band light source. The simulations also indicate that a carrier loss term that depends on the resonant laser power is present. The simulations are predictive, yet consider no effects that can explain the reduction in emission linewidth. Thus, another process must be present and we discuss likely possibilities.

\vspace{0.5cm}\noindent\textbf{Experiments.} The sample was made by molecular beam epitaxy.
It consists of strain-induced InAs QDs in GaAs embedded in a 4$\lambda/n$ thick planar distributed Bragg reflector cavity ($\lambda$ is the cavity resonance wavelength and $n$ is the GaAs refractive index). The cavity can enhance the fluorescence collection through the upper sample surface resulting in extraction efficiencies of $10~\%-20~\%$~\cite{Benisty1998}. We observe a single charged exciton and show in Fig.~1(c) that the observed emission has a low multi-photon contribution: $g^{(2)}(0) = 0.22 \pm 0.03$. From data in Fig.~1(c), we determine the radiative decay rate, {$\Gamma = 1.5~\text{ns}^{-1}$, and thus a radiative-limited linewidth of 250~MHz.

To suppress the resonant laser scattering from the QD fluorescence signal, we use an orthogonal excitation-collection scheme: the excitation laser light is fiber-coupled to the side of the sample and propagates in-plane and the collection axis is normal to the plane (Fig.~\ref{fig:Schematics}(d))~\cite{Muller2007,Muller2008}. The in-plane pump laser couples predominantly to the $4\lambda/n$ cavity region which forms an in-plane waveguide. The QD emission is  collected vertically and is coupled into a single-mode fiber. 

The resonant laser light is a tunable continuous-wave~(CW) laser (200~kHz linewidth). Its frequency is measured using a wavelength meter. The weak above-band light is a 633~nm HeNe laser and is sent to the sample surface through the collection optical fiber. Polarizers and polarization controllers inserted in the excitation laser path and collection path further minimize the detected laser scattering. The sample is cooled to $4.2$~K using a helium bath cryostat. 

As has been previously reported~\cite{Metcalfe2010,Nguyen2012}, the  on-resonance fluorescence signal is weak for most QD states, and weak above-band light can enhance the fluorescence intensity  (Fig.~\ref{fig:Schematics}(a)). We measure large photon intensity improvements for many QDs, but not for all of them. For all the results of this letter, the power of the HeNe laser beam on the sample is on the order of the nano-watts and no photoluminescence (PL) signal from the HeNe alone is measured. Measurements were made at a variety of Rabi frequencies varying from near zero to 0.7~GHz (Fig.~1(c)).

\vspace{0.5cm}\noindent\textbf{Resonance fluorescence maps.} To investigate the effects of the weak HeNe laser light on the emission properties from exciton transitions in resonant excitation, we measure resonance fluorescence maps. Such maps represent the QD fluorescence intensity, $I_{\text{Exp}}$ as a function of both the pump laser frequency, $\omega_{\text{L}}$ and the fluorescence photons frequency, $\omega_{\text{Ph}}$. To do so, we use the tunable CW laser and a 200~MHz band-pass Fabry-P\'erot filter to measure the emitted photons frequency. We obtain a map through a series of PL spectroscopy scans (i.e.~by scanning the Fabry-P\'erot filter frequency) for several laser frequencies. An identical map could also be obtained by scanning the laser frequency and keeping the filter position constant for each scan. We measure the QD fluorescence intensity for each point using a single-photon avalanche photodiode connected to a photon-counting module.

We use the detuning-dependent resonant fluorescence equations of a coherently driven two-level system~\cite{Mollow1969,Ulhaq2013} to calculate fluorescence maps  in different scattering regimes. Results in the inelastic scattering regimes are plotted and discussed in the Supplemental Material. We discuss here the elastic scattering regime (also called the Rayleigh scattering regime). In the ideal case (narrow linewidth laser, Fabry P\'erot filter and QD state), the map is a point centered on the QD state frequency. For a real world case, the ideal map  is successively convolved by the laser lineshape (convolved along the $\omega_L$ axis), the detector response (convolved along the $\omega_{\text{Ph}}$ axis) and the QD state broadening (convolved along the diagonal direction, $\omega_{\text{ph}}=\omega_{\text{L}})$. The QD state broadening is Gaussian in the case of inhomogeneous broadening and Lorentzian in the case of homogeneous broadening~\cite{Loudon2000}. Thus, one can distinguish between inhomogeneous and homogeneous broadening by looking at the diagonal lineshape. The calculated maps that we obtain are plotted in \mbox{Figs.~\ref{fig:maps}(a)-(b)}.

\begin{figure}[t!] 
\includegraphics[trim=0cm 0cm 0cm 0cm, clip=true,width=1\linewidth]{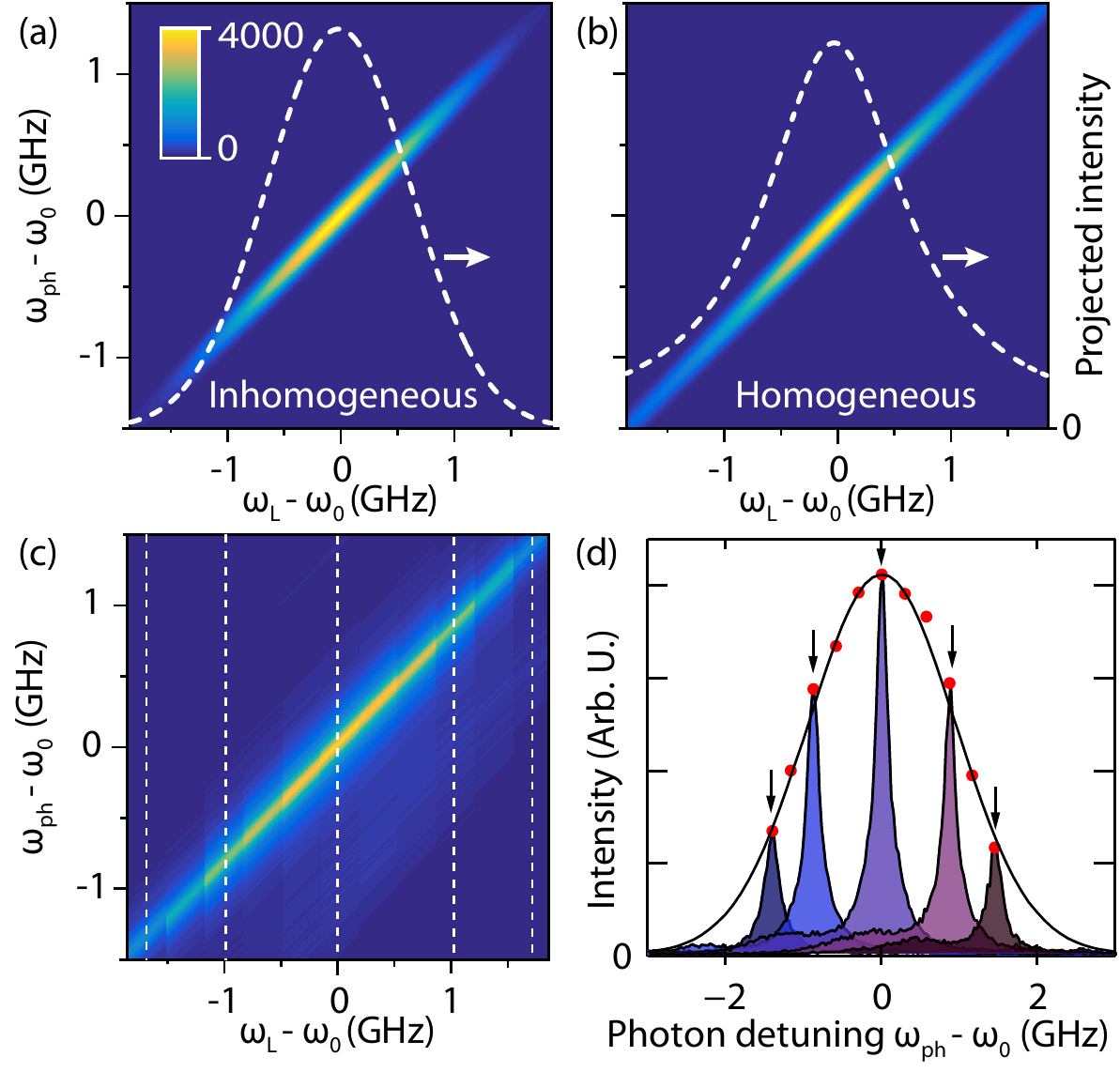}
\caption{(a-b) Theoretical fluorescence maps of emission from a QD transition with inhomogeneous (a) or homogeneous broadening~(b). The shape of the intensity distribution in the diagonal $\omega_{\text{ph}}=\omega_{\text{L}}$ is Gaussian in the inhomogeneous case and Lorentzian in the homogeneous case as indicated by the white dotted lines. (c) Experimental fluorescence map obtained by sweeping the frequency of a laser, $\omega_{\text{L}}$ and of a Fabry P\'erot filter, $\omega_{\text{ph}}$ around the QD frequency, $\omega_0$. (d) Shaded curves: PL spectra obtained for different laser frequencies, as indicated by dotted lines in (c).  The red dots indicate the PL maximum for each laser frequency. The envelope intensity is Gaussian (black line).}
\label{fig:maps}
\end{figure}

A measured fluorescence map $I_{\text{Exp}}(\omega_{\text{Ph}},\omega_{\text{L}})$ is shown in Fig.~\ref{fig:maps}(c) in the elastic scattering regime with the additional HeNe source ($\approx$ 20 nW). We observe that the fluorescence map has an oval-like shape strongly elongated along the $\omega_{\text{L}}=\omega_{\text{Ph}}$ diagonal. From this fluorescence map, we take PL spectra for specific values of $\omega_{\text{L}}$ (vertical dotted lines in Fig.~\ref{fig:maps}(c)). Example PL spectra are plotted in Fig.~\ref{fig:maps}(d) for these laser frequencies. Each spectrum has a Lorentzian line-shape which is determined by a convolution of the laser line and the Fabry-P\'erot interferometer response. 
Each spectrum has a 200~MHz Lorentzian linewidth limited by the resolution of our Fabry-P\'erot interferometer. 
Red dots in Fig.~\ref{fig:maps}(d) represent the PL peak amplitudes for different laser frequencies in the fluorescence map. Here, the amplitude of the spectra corresponds to the probability of overlap of the broadened QD transition frequency with the laser frequency. 
We observe that the amplitude of the PL spectrum as a function of the laser detuning is fit by a Gaussian distribution function (solid black line) and thus deduce that the broadening is inhomogeneous. This means that the QD transition energy is not constant in time but varies. The full width at half maximum of the Gaussian distribution of the emission is $\approx$~2.5~GHz in Fig.~\ref{fig:maps}(d).
Using second order autocorrelation measurements, we previously reported energy jitter times of $\approx24~$ns with a similar sample~\cite{Thomay2016}. 

\vspace{0.5cm}\noindent\textbf{\bf Linewidth narrowing and enhanced intensity.} To investigate the effect of the above-band laser power on the resonant emission, we perform a series of near resonance photoluminescence excitation (PLE) spectroscopy scans. In each PLE spectrum we sweep the CW laser through the QD transition and count detected QD photons as a function of the laser frequency. The PLE lineshape can be written as a convolution of the Lorentzian (homogeneous) and the Gaussian (inhomogeneous) components of the QD emission, leading to a Voigt function. The Lorentzian component is due to the radiative lifetime and any power broadening induced by the laser. Its full width at half maximum is determined by the relation $\sqrt{T_1^{-2}+2\Omega_r^2}$ for each resonant laser power where $\Omega_r$  is the Rabi frequency~\cite{Flagg2009a}. $\Omega_r$ is determined using continuous wave second-order autocorrelations on the QD photons~\cite{Flagg2009a}, shown in Fig~1(c). We fit the PLE spectra using such a function to extract the amplitude and the linewidth of the Gaussian component.

Using the PLE fits described above, we plot in Fig.~\ref{fig:complete}(a) the PLE intensity integrated over a resonant laser frequency sweep, and in Fig.~\ref{fig:complete}(b) the inhomogeneous linewidth, both as a function of the HeNe power. These figures show that adding the additional above band laser does not only increase the number of emitted photons, but it also introduces a linewidth narrowing by a factor of about~2. While the linewidth decreases and the peak intensity increases, the integrated intensity is not conserved.  Further, the amount of above-band light needed to maximize the brightness is a function of resonant excitation laser power. The maximum of intensity, visible in Fig.~\ref{fig:complete}(a), and the minimum linewidth, in Fig.~\ref{fig:complete}(b), shift towards higher HeNe powers when the resonant laser power increases.

To gain more insight into the dynamics of the resonant excitation with additional HeNe laser, we programmed a Monte Carlo simulation to model the emission dynamics. The model is based on probabilities of events occurring in a single particle picture. It considers separate electron~($\mathrm{e^-}$) and hole~($\mathrm{h^+}$) empty and occupied ground states, with fermion statistics. A higher excited state for each charged carrier (with infinite occupation) can also be populated.

We simulate all the experiments with a charged exciton and a single parameter set, taking only the five most crucial parameters into account (see solid lines in Fig.~\ref{fig:complete}(a)). They are a radiative decay of $\mathrm{e^-}$ and $\mathrm{h^+}$ with opposite spin, creation and capture of charges by the above-band light, resonant excitation of $\mathrm{e^--h^+}$ pairs, non-radiative relaxation from the excited state to the ground state, and a resonant laser induced charge carrier loss term. We also consider radiative decay from the excited state, spin flip of a charge, asymmetric capture of carriers, and loss of charges only from the excited level. Nevertheless, these parameters are not needed. To test the model we simulated the PL of an empty QD with and without resonant excitation. Both produce the expected result for a very broad range of parameters, i.e. the resonant excitation creates excitons, but no charged excitons, if the QD is empty. For example, it models the power dependence of the above-band laser in the high-power regime in Fig.~3(a) and beyond. It also can reproduce the behavior of the data measured in Ref.~\cite{Nguyen2012}, but an asymmetric flux of $\mathrm{h^+}$ and $\mathrm{e^-}$ is needed, as also claimed by the authors in Ref.~\cite{Nguyen2012}.

Fig.~\ref{fig:complete}(b) shows that the linewidth narrowing is also a function of HeNe power and resonant laser power. For each resonant laser power the linewidth narrows with increased HeNe power. For higher resonant laser power the linewidth narrows more slowly. The linewidth does not narrow to the radiative linewidth limit (250~MHz) but levels off at just below~1.4~GHz.

\begin{figure}[t]
\includegraphics[trim=0cm 0.5cm 0cm 0cm, clip=true, width=1\linewidth]{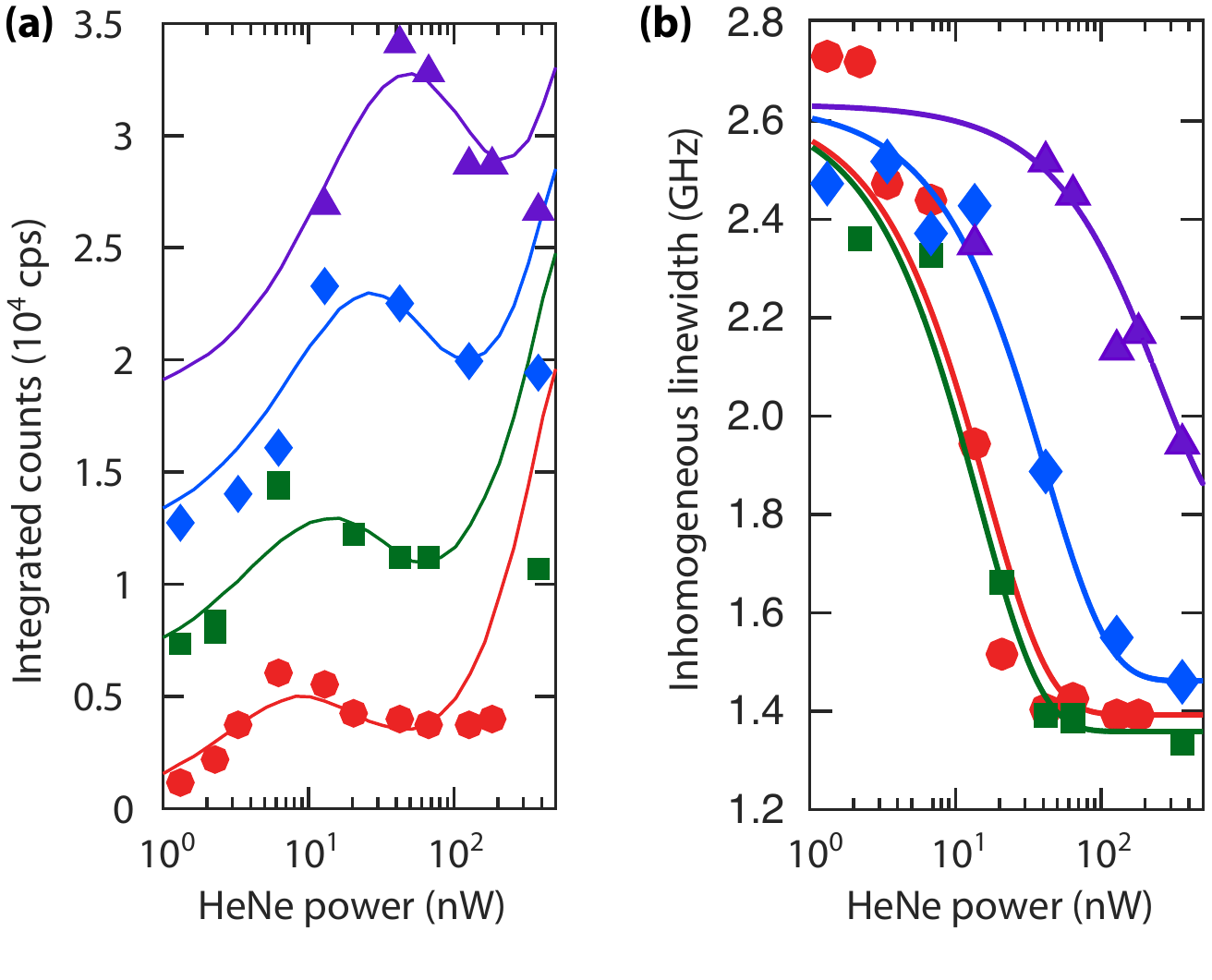}
\vspace{-0.5cm}
\caption{\label{fig:complete} The different sets of lines are due to different resonant laser powers. Red circles: $\Omega_r=0.23~$GHz, green squares: $\Omega_r=0.30~$GHz, blue diamonds: $\Omega_r=0.47~$GHz, violet triangles: $\Omega_r=0.66~$GHz). (a) Observed integrated intensity as a function of HeNe power. The solid lines represent the simulation. The data with $\Omega_r=0.47$ GHz (blue diamonds) has an increase of integrated brightness of 17 with respect to its lowest value. The four curves were offset by 6000 counts each for clarity. The increase in counts on the right end of the simulation is due to photoluminescence created directly by the HeNe laser. (b) The QD inhomogeneous linewidth decreases with increased HeNe power. The decrease is slower for higher resonant laser powers. The solid lines represent exponential~fits.}
\vspace{-0.2cm}
\end{figure}

{\bf Discussion.}
The Monte Carlo simulations show that if additional carriers are provided to the QD, the time the QD spends in the desired exciton state can be extended, increasing the photon flux. An important component of the model is a charge loss term that depends on the resonant-laser power. This loss term implies that additional carriers are needed to stabilized the QD population, and these carriers are provided by the above-band light source. However, if the above-band light creates too many charges, the total amount of emitted photons will decrease again. Since the model is rate-based it cannot predict any change in linewidth, suggesting another process contributes to the inhomogeneous broadening changes.

An exponential dependent change in the inhomogeneous exciton emission linewidth with above-band laser power is observed for over two decades of laser power. The source of the linewidth narrowing is not covered by our model and thus, discussion of its causes is only speculative. Two processes associated with additional carriers in the vicinity of the QD could explain the linewidth change. Screening of fluctuating charges and trap states with the addition of charged carriers can result in motional narrowing~\cite{Bjorkstam1985} and will be exponential in carrier population. Such traps could be local wetting layer fluctuations, nearby QD states, or defects \cite{Houel2012}. Additionally, stabilizing~\cite{Nguyen2013}, as opposed to screening, fluctuating traps populations with additional carriers will have a Poissonian cumulative distribution function with a similar dependence. From our data we cannot distinguish between these two effects and thus cannot conclude if either is responsible for the observed effect.

Our data also shows that the linewidth narrowing depends on the resonant laser power. While the initial exciton linewidth is not a function of Rabi frequency, additional above-band laser power is required as the Rabi frequency increases to obtain the same reduction in linewidth. This could be linked to the resonant laser induced charge loss term in the Monte Carlo simulation that is responsible for the shift of the brightness maximum. A higher carrier loss caused by the resonant laser from the QD to the local environment could require additional above-band light to be neutralized. 

What is clear from the Monte Carlo simulations is that the resonant laser effects the carrier population inside the QD. Neutralizing this loss with carriers from the above-band laser increases the resonant fluorescence. Voltage biased devices have been used in several QD experiments~\cite{Finley2004,Houel2012}, and have been recently used in QD single-photon sources~\cite{Somaschi2015,Ding2016a}. While these structures should aid in stabilizing the local charge environment around the QD, they likely will not neutralize the resonant laser induced loss from the QD \cite{Kurzmann2016a,Kurzmann2016}.

In conclusion, we have experimentally investigated the radiative emission from single QD states under resonant excitation. To study these effects we have introduced fluorescence maps produced by varying the detection frequency in subsequent PL scans near the emission resonance. We show the QD transition broadening is largely inhomogeneous. Using a weak above-band light, the QD peak emission increases by over 30 times, where the integrated intensity increases 17 times. Simultaneously, the inhomogeneous emission linewidth decreases by a factor of about~2. The change in total integrated intensity is well-modeled by a simple Monte Carlo simulation where a loss term that depends on the resonant laser power must be included.

\section*{Acknowledgments}
This work is partially supported by Physics Frontier Center at the Joint Quantum Institute (PFC@JQI) and the Army Research Laboratory.

\bibliography{Articles-biblio-PaperFluoReso2016}





\clearpage

\onecolumngrid
\vspace{\columnsep}
\begin{center}
\textbf{\large Supplemental material: Quantum dot resonant-fluorescence linewidth narrowing and enhanced intensity}
\end{center}
\vspace{\columnsep}
\twocolumngrid

\setcounter{equation}{0}
\setcounter{figure}{0}
\setcounter{table}{0}
\makeatletter
\renewcommand{\theequation}{S\arabic{equation}}
\renewcommand{\thefigure}{S\arabic{figure}}

\section{Fluorescence map in the inelastic scattering regime}

We use the detuning-dependent resonant fluorescence equations of a coherently driven two-level system~\cite{Mollow1969,Ulhaq2013} to calculate fluorescence maps in the inelastic scattering regime. Maps without broadening are plotted in Fig.~S1(a),(c),(f). In the low pump power regime ($\Omega_r<\Gamma$), the maps show a small dot centered on the QD transition frequency, $\omega_0$ (Fig. S1(a),(c)). In the strong power regime ($\Omega_r>\Gamma$), Mollow Triplet branches appear, as visible in Fig. S1(f). 

When some broadening effects are introduced in the inelastic scattering regime, the maps are modified. For homogeneous broadening, the map has a diamond shape as shown in Fig.~\ref{fig:supp1}(b), aligned along the horizontal (constant value of $\omega_{\text{ph}}$) and vertical axis (constant value of~$\omega_{\text{L}}$) of the fluorescence map. The fluorescence map is not circular because the fluorescence probability is reduced for off axis points: the laser and the observed photon are detuned from the center of the transition. Thus, for pure dephasing the diamond is larger than the natural radiative linewidth but the diamond shape and orientation remains the same because the central QD transition energy is unchanged by pure dephasing. 

Inhomogeneous broadening is different as it modifies the QD transition energy over a frequency range $\omega_G$: the QD transition energy is now time dependent but can be considered constant during the emission process. The laser acts as a narrow band-pass filter since emission from the QD transition is only observed when the QD transition sweeps through the laser. The inhomoheneous broadening of the transition corresponds to a convolution of the ideal fluorescence map with a Gaussian in the diagonal axis ($\omega_{\text{ph}}=\omega_{\text{L}}$). Calculated maps are plotted in Figs. S1(c)-(h) for several Rabi frequencies and inhomogeneous broadenings.

If the system was in the Rayleigh (e.g. elastic) scattering regime, the map would have an oval-like shape and would be elongated along the diagonal axis ($\omega_{\text{ph}}=\omega_{\text{L}}$) as described in the main text and shown in Figs. 2(a),(b).

\begin{figure*}
\includegraphics[width=0.75\linewidth]{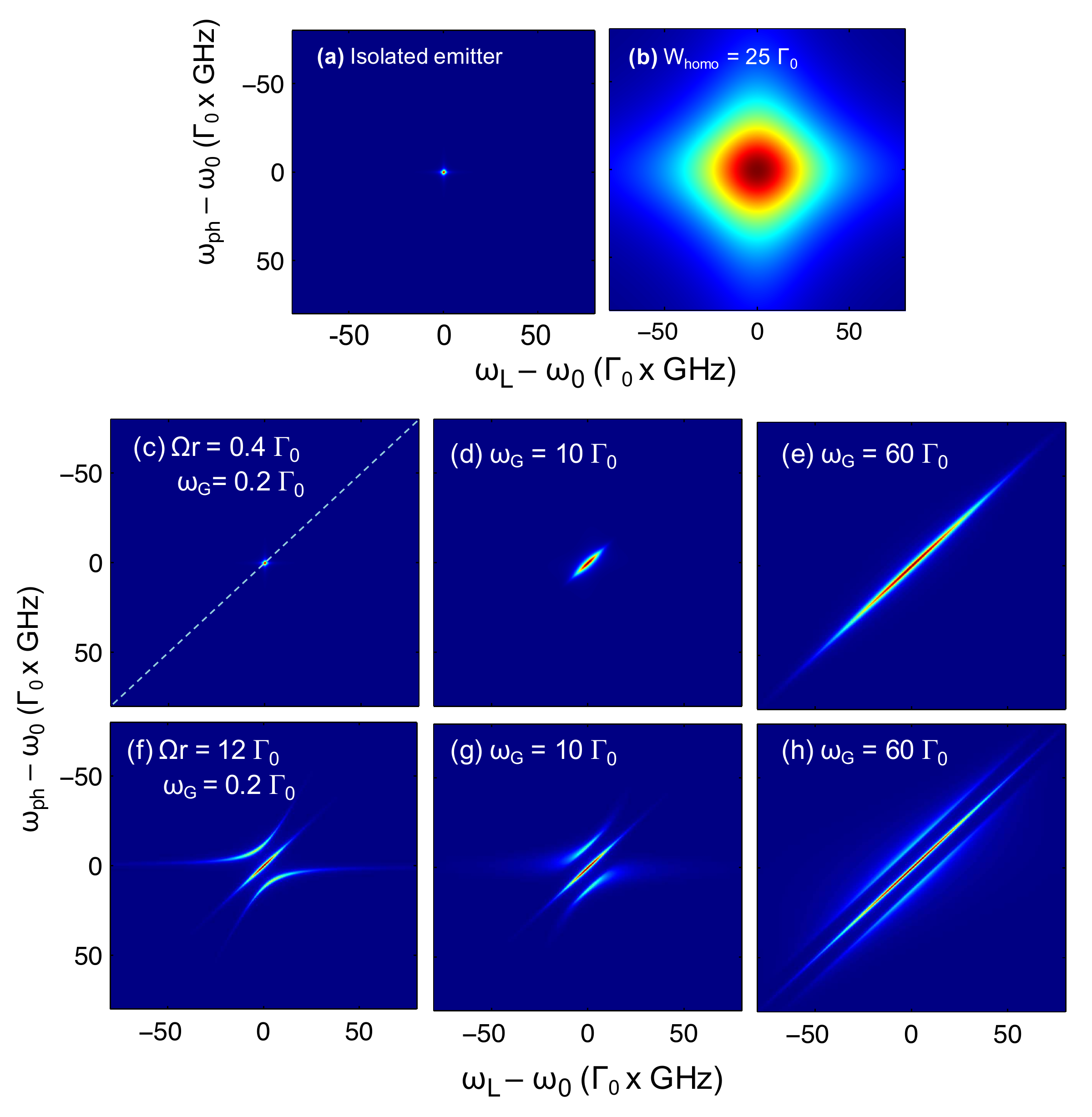}
\caption{Theoretical fluorescence maps in the inelastic scattering regime. (a,c,f) Case of an ideal emitter in the low pump power regime (a,c) and strong power regime (f). (b) With homogeneous broadening the maps has a diamond shape. (d,e,g,h) With inhomogeneous broadening, the maps are convoluted by a Gaussian in the diagonal axis to account for the spectral wandering of the emitter transition energy. For all the figures, the detector and the laser are supposed to be spectrally much narrower than the QD transition linewidth, $\Gamma_0$.}
\label{fig:supp1}
\end{figure*}


\end{document}